\documentstyle[12pt,axodraw]{article}

\catcode`@=12
\begin{document}
\thispagestyle{empty}
\begin{flushright} 
UCRHEP-T304\\ 
April 2001\
\end{flushright}
\vspace{0.5in}
\begin{center}
{\large	\bf Global U(1) Invariance and Mass Scales in Supersymmetry\\}
\vspace{1.5in}
{\bf Ernest Ma\\}
\vspace{0.2in}
{\sl Physics Department, University of California, Riverside, 
California 92521\\}
\vspace{1.5in}
\end{center}
\begin{abstract}\
In a supersymmetric theory with global U(1) invariance, the spontaneous 
breaking of the latter without the breaking of supersymmetry is revisited 
in the case of the most general superpotential of 3 singlet superfields.  
The interesting possibility of having 2 hierarchical mass scales is pointed 
out, together with its consequences as applied to the axionic solution of 
the strong CP problem.
\end{abstract}

\newpage
\baselineskip 24pt

\section{Introduction}

The spontaneous breaking of a continuous global symmetry is well-known 
\cite{spont} to result in a massless Nambu-Goldstone boson.  If this 
happens in a supersymmetric theory \cite{wz} without the breaking of the 
supersymmetry, a massless superfield must emerge, resulting in the 
existence of ``flat directions'' \cite{flat}.  If the supersymmetry is 
also broken, then the components of this superfield will become 
massive, except for the Nambu-Goldstone boson.  The only question is: 
how is the scale of supersymmetry breaking related to the scale of 
spontaneous breaking of the continuous global symmetry?  Naively, we 
would expect them to be the same.  On the other hand, it is desirable in 
the implementation of the axionic solution \cite{pq,ww} of the strong CP 
problem \cite{cpv} that the $U(1)_{PQ}$ symmetry breaking scale, presumably 
of order $10^{9}$ to $10^{12}$ GeV, be much greater than the supersymmetry 
breaking scale, presumably of order 1 TeV.

In this paper, this problem is revisited in the case of the most general 
superpotential of 3 singlet superfields.  The relationship of the 2 
arbitrary mass scales of this superpotential to the $U(1)$ breaking scale 
is clarified, particularly with respect to a simplified form \cite{form} 
which is widely used in the literature.  The possibility of having a 
``seesaw'' mass spectrum with new bosons and fermions at the TeV scale 
is pointed out.

\section{Superpotential of 3 Singlet Superfields}

Consider 3 singlet superfields $\hat \phi_1$, $\hat \phi_2$, and $\hat \chi$, 
transforming as $+1$, $-1$, and 0 under a global $U(1)$ symmetry.  The 
following simple superpotential,
\begin{equation}
\hat W = f \hat \chi (\hat \phi_1 \hat \phi_2 - \Lambda),
\end{equation}
is invariant under $U(1)$ and is widely used in axionic supersymmetric 
models.  The idea is that the supersymmetric minimum of the corresponding 
scalar potential is given by
\begin{equation}
v_1 v_2 = \Lambda,
\end{equation}
where $v_{1,2} = \langle \phi_{1,2} \rangle$, which breaks $U(1)$ 
spontaneously and gives rise to the axion.  However, it is clear that 
Eq.~(1) is missing the allowed term $\mu_{12} \hat \phi_1 \hat \phi_2$ which 
must be set equal to zero by hand to obtain Eq.~(2).  Nevertheless, as 
long as the supersymmetry is exact, the condition $\mu_{12} = 0$ is maintained 
to all orders in perturbation theory.  In reality, we know the supersymmetry 
is broken, hence a natural lower bound for $\mu_{12}$ is $M_{SUSY} \sim 1$ TeV.
This has the important phenomenological consequence that all particles 
associated with the axion (including the axino) must not be much lighter 
than $M_{SUSY}$.

Instead of the simplified form of Eq,~(1), consider the most general 
superpotential of $\hat \phi_1$, $\hat \phi_2$, and $\hat \chi$, i.e.
\begin{equation}
\hat W = {1 \over 2} m \hat \chi^2 + {1 \over 3} h \hat \chi^3 + 
\mu \hat \phi_1 \hat \phi_2 + f \hat \chi \hat \phi_1 \hat \phi_2.
\end{equation}
Under the transformation $\hat \chi \to \hat \chi + u$, we have
\begin{equation}
\hat W \to {1 \over 2} m u^2 + {1 \over 3} h u^3 + u(m+hu) \hat \chi 
+ {1 \over 2} (m + 2 h u) \hat \chi^2 + {1 \over 3} h \hat \chi^3 + 
(\mu + f u) \hat \phi_1 \hat \phi_2 + f \hat \chi \hat \phi_1 \hat \phi_2.
\end{equation}
If we now require
\begin{equation}
\mu+fu = 0, ~~~ f \Lambda = -u(m + h u),
\end{equation}
then after dropping the constant term, we have
\begin{equation}
\hat W = {1 \over 2} (m + 2 h u) \hat \chi^2 + {1 \over 3} h \hat \chi^3 
+ f \hat \chi (\hat \phi_1 \hat \phi_2 - \Lambda),
\end{equation}
which becomes Eq.~(1) in the limit $m \to 0$, $h \to 0$, but $\Lambda$ 
remains finite.  Thus the \underline {hidden assumptions} of Eq.~(1) are 
that $m$ is very small, but $u$ (hence $\mu$) is 
very large, such that $\Lambda$ is the order of their product.  Since $m$ 
should be bounded from below by $M_{SUSY}$ and $\mu$ from above by 
$M_{Planck}$, a reasonable value for $\Lambda$ is indeed $10^3$ GeV $\times$ 
$10^{19}$ GeV = ($10^{11}$ GeV)$^2$ and suitable for the axion scale.

Consider now Eq.~(6) with $m \neq 0$ and $h \neq 0$.  The two fundamental 
mass scales $m$ and $\mu$ of Eq.~(3) may be chosen arbitrarily and 
independently as long as each is greater than $M_{SUSY}$.  However, the 
choice $m \sim M_{Planck}$ and $\mu \sim M_{SUSY}$, i.e. opposite to that of 
Eq.~(1), is very natural and has interesting consequences, to be shown below.

\section{Breaking of the U(1) Symmetry}

Let the superfields $\hat \phi_1$ and $\hat \phi_2$ be redefined as $v_1 + 
\hat \phi_1$ and $v_2 + \hat \phi_2$, with $v_1 v_2 = \Lambda$, then
\begin{equation}
\hat W = {1 \over 2} (m + 2 h u) \hat \chi^2 + {1 \over 3} h \hat \chi^3 
+ f \hat \chi (v_2 \hat \phi_1 + v_1 \hat \phi_2 + \hat \phi_1 
\hat \phi_2),
\end{equation}
which shows that the superfields
\begin{equation}
\hat \chi, ~~~ \hat \eta \equiv {v_2 \hat \phi_1 + v_1 \hat \phi_2 \over 
\sqrt {v_1^2 + v_2^2}},
\end{equation}
are massive with a mass matrix given by
\begin{equation}
{\cal M} = \left[ \begin{array} {c@{\quad}c} m + 2 h u & f v \\ f v & 0 
\end{array} \right],
\end{equation}
where $v = \sqrt {v_1^2 + v_2^2}$ and that the superfield
\begin{equation}
\hat \zeta \equiv {v_1 \hat \phi_1 - v_2 \hat \phi_2 \over \sqrt {v_1^2 + 
v_2^2}}
\end{equation}
is massless.  In terms of $\hat \chi$, $\hat \eta$, and $\hat \zeta$,
\begin{equation}
\hat W = {1 \over 2} (m + 2 h u) \hat \chi^2 + {1 \over 3} h \hat \chi^3 + 
f v \hat \chi \hat \eta + scf \hat \chi (\hat \eta \hat \eta - 
\hat \zeta \hat \zeta) + (c^2 - s^2) f \hat \chi \hat \eta \hat \zeta,
\end{equation}
where $s = \sin \theta$, $c = \cos \theta$, and $\tan \theta = v_2/v_1$.

In anticipation of the assumed symmetric soft breaking of the supersymmetry 
which will set $v_1 = v_2$, we simplify Eq.~(11) by taking $s = c = 
1/\sqrt 2$.  In addition, using $v^2 = 2 \Lambda$ and Eq.~(5), we assume 
$|\mu| << |m|$.  Hence
\begin{equation}
fv = f \sqrt {2 \Lambda} \simeq \sqrt {2 \mu m} << |m|
\end{equation}
in Eq.~(9).  Thus the heavy superfield $\hat \chi$ can be integrated out 
and the effective $\hat \eta$ acquires a ``seesaw'' mass, i.e.
\begin{equation}
m_\eta \simeq -{f^2 v^2 \over m} \simeq - 2 \mu.
\end{equation}
The effective superpotential of $\hat \eta$ and $\hat \zeta$ is then given by
\begin{equation}
\hat W = {1 \over 2} m_\eta \hat \eta \hat \eta - {m_\eta \over 2 v} 
\hat \eta (\hat \eta \hat \eta - \hat \zeta \hat \zeta).
\end{equation}

The corresponding contribution to the Lagrangian is
\begin{eqnarray}
-{\cal L}_{int} &=& |m_\eta \eta - (3 m_\eta / 2 v) \eta^2 + (m_\eta 
/ 2 v) \zeta^2|^2 + |(m_\eta / v) \eta \zeta|^2 \nonumber \\ 
&+& [(m_\eta/2) \tilde \eta \tilde \eta - (3 m_\eta / 2 v) \eta \tilde \eta 
\tilde \eta + (m_\eta / 2 v) \eta \tilde \zeta \tilde \zeta + (m_\eta / v) 
\zeta \tilde \zeta \tilde \eta + h.c.]
\end{eqnarray}

Since the scalar field $\eta$ and the fermion field $\tilde \eta$ are 
massive, they can be integrated out to obtain the effective Lagrangian of 
$\zeta$ and $\tilde \zeta$.  To lowest order, it should be 
of the form
\begin{equation}
-{\cal L}_{eff} = a |\zeta|^2 + b |\zeta|^4 + c \tilde \zeta \tilde \zeta.
\end{equation}
At tree level, it is clear that $a = c = 0$, and $b$ is given by the 
diagrams of Fig.~1, i.e.
\begin{equation}
b = \left( {m_\eta \over 2 v} \right)^2 - {(m_\eta^2/2v)^2 \over m_\eta^2} = 0,
\end{equation}
which is exactly as expected.  In fact, $a=b=c=0$ is true to all orders in 
perturbation theory as long as the supersymmetry is not broken because 
$\hat \zeta$ is a massless superfield and the symmetry of Eq.~(14) forbids 
the $\hat \zeta^3$ term in $\hat W_{eff}$.

\section{Breaking of the Supersymmetry}

To break the supersymmetry softly, we go back to Eq.~(3) and consider
\begin{equation}
V_{soft} = \mu_\chi^2 |\chi|^2 + \mu_0^2 (|\phi_1|^2 + \phi_2|^2) + 
[B \mu \phi_1 \phi_2 + A f \chi \phi_1 \phi_2 + h.c.]
\end{equation}
The minimum of the complete scalar potential is then given by
\begin{eqnarray}
V_{min} &=& (m u + h u^2 + f v_1 v_2)^2 + (\mu + f u)^2 (v_1^2 + v_2^2) 
\nonumber \\ &+& \mu_\chi^2 u^2 + \mu_0^2 (v_1^2 + v_2^2) + 2 B \mu v_1 v_2 
+ 2 A f u v_1 v_2.
\end{eqnarray}
Using $v_1 = v_2 = v/\sqrt 2$, and $|\mu| << |m|$, we find
\begin{equation}
v^2 \simeq -{2 m u \over f},
\end{equation}
with $u$ given by
\begin{equation}
3 f^2 u^2 + 2 f (2 \mu + A) u + \mu^2 + \mu_0^2 + B \mu = 0.
\end{equation}
In the limit $V_{soft} = 0$, we recover the supersymmetric solution of Eq.~(5) 
as expected.  Now if we set $\mu \sim M_{SUSY}$ as well as $\mu_0$, $B$, and 
$A$, then $u \sim M_{SUSY}$ is assured.

We can again use the seesaw approximation to integrate out $\chi$ and 
obtain the masses of $\eta$ and $\zeta$.  Instead of $m_\eta \simeq -2 \mu$ 
and $m_\zeta = 0$ in the supersymmetric limit, we obtain
\begin{eqnarray}
&& m_{Re \eta}^2 \simeq 2 \mu [2 \mu - 4 (\mu + f u) + 2 A], \\ 
&& m_{Im \eta}^2 \simeq 2 \mu [2 \mu + 2 A] + 2 (\mu + f u )^2 + 2 \mu_0^2, \\ 
&& m_{Re \zeta}^2 = 2 (\mu + f u)^2 + 2 \mu_0^2, \\ 
&& m_{Im \zeta}^2 = 0.
\end{eqnarray}
Since $\mu_{12} \tilde \phi_1 \tilde \phi_2$ is an allowed soft supersymmetry 
breaking term, the mass of the fermion $\tilde \eta$ is now $|-2 \mu + 
\mu_{12}|$ and that of $\tilde \zeta$ is $|\mu_{12}|$.

\section{Physical Consequences}

Starting with Eq.~(3), we have shown that there are 2 interesting limits. 
If $m \to 0$ and $\mu \to \infty$ with $v \sim \sqrt {m \mu}$ finite, we 
get Eq.~(1) with the following particle spectrum.  The superfields $\hat \chi$ 
and $\hat \eta$ are heavy with mass $fv$.  The scalar field $\zeta$ 
contains the axion as its phase, with $v$ as the scale of $U(1)$ symmetry 
breaking, but the mass of $|\zeta|$ is the order of $M_{SUSY}$ as is the 
mass of the axino $\tilde \zeta$.  At the other extreme with $m \sim 
M_{Planck}$ and $\mu \sim M_{SUSY}$, we get also $v \sim \sqrt {m \mu}$ which 
is now fixed at $\sim 10^{11}$ GeV, and suitable for the axion scale.  The 
particle spectrum in this case is different.  The superfield $\hat \chi$ is 
heavy with mass $m$, but both $\eta$ and $|\zeta|$, as well as $\tilde \eta$ 
and $\tilde \zeta$ have masses the order of $M_{SUSY}$.

\section{Concluding Remarks}

In general, the values of $m$ and $\mu$ are arbitrary and independent, and 
protected against large radiative corrections by the supersymmetry.  So as 
long as they are greater than $M_{SUSY}$, the field theory is well-behaved 
and the $U(1)$ breaking scale of $\sqrt {m \mu}$ is generated.  If only one 
mass scale is allowed by the choice of $U(1)$ charge assignments as in a 
recent axionic supersymmetric model for neutrino masses \cite{ma01}, then 
this mechanism is not available, resulting in the breaking of $U(1)$ at 
that one mass scale \cite{more}.  Nevertheless, $M_{SUSY} << v$ is possible in all 
these examples.

\section*{Acknowledgement}

This work was supported in part by the U.~S.~Department of Energy
under Grant No.~DE-FG03-94ER40837.

\newpage
\bibliographystyle{unsrt}

\begin{thebibliography}{99}
\bibitem{spont} Y. Nambu, Phys. Rev. Lett. {\bf 4}, 380 (1960); J. Goldstone, 
Nuovo Cimento {\bf 19}, 154 (1961); J. Goldstone, A. Salam, and S. Weinberg, 
Phys. Rev. {\bf 127}, 965 (1962).
\bibitem{wz} J. Wess and B. Zumino, Nucl. Phys. {\bf B70}, 34 (1974).
\bibitem{flat} I. Affleck, M. Dine, and N. Seiberg, Nucl. Phys. {\bf B256}, 
557 (1985); G. Costa, F. Feruglio, F. Gabbiani, and F. Zwirner, Nucl. Phys. 
{\bf B286}, 325 (1987); E. Ma, Mod. Phys. Lett. {\bf A14}, 1637 (1999).
\bibitem{pq} R. D. Peccei and H. R. Quinn, Phys. Rev. Lett. {\bf 38}, 1440 
(1977).
\bibitem{ww} S. Weinberg, Phys. Rev. Lett. {\bf 40}, 223 (1978); F. Wilczek, 
Phys. Rev. Lett. {\bf 40}, 279 (1978).
\bibitem{cpv} C. G. Callan, R. F. Dashen, and D. J. Gross, Phys. Lett. 
{\bf B63}, 334 (1976); R. Jackiw and C. Rebbi, Phys. Rev. Lett. {\bf 37}, 172 
(1976).
\bibitem{form} J. E. Kim, Phys. Lett. {\bf B136}, 378 (1984).
\bibitem{ma01} E. Ma, hep-ph/0102008.
\bibitem{more} Using 4 singlet superfields with PQ charge assignments +2, 0, 
--1, --2, we can again have 2 mass scales $m$ and $\mu$.  The axion scale 
of $\sqrt {m \mu}$ is then generated and can be identified with the mass 
scale of the heavy singlet neutrino as in Ref.[8].
\end{thebibliography}

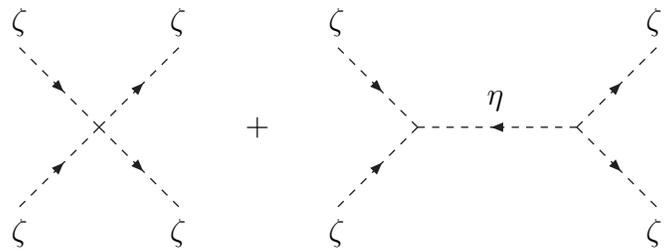
\begin{figure}
\begin{center}
\begin{picture}(270,110)(0,0)
\DashArrowLine(10,20)(40,50){3}
\DashArrowLine(10,80)(40,50){3}
\DashArrowLine(40,50)(70,20){3}
\DashArrowLine(40,50)(70,80){3}
\Text(100,50)[]{+}
\DashArrowLine(130,20)(160,50){3}
\DashArrowLine(130,80)(160,50){3}
\DashArrowLine(220,50)(160,50){3}
\DashArrowLine(220,50)(250,20){3}
\DashArrowLine(220,50)(250,80){3}
\Text(10,10)[]{$\zeta$}
\Text(70,10)[]{$\zeta$}
\Text(130,10)[]{$\zeta$}
\Text(250,10)[]{$\zeta$}
\Text(10,90)[]{$\zeta$}
\Text(70,90)[]{$\zeta$}
\Text(130,90)[]{$\zeta$}
\Text(250,90)[]{$\zeta$}
\Text(190,60)[]{$\eta$}

\end{picture}
\end{center}

\caption{Contributions to the effective quartic coupling $|\zeta|^4$.}
\end{figure}

\end{document}